\documentclass{elsart}
\usepackage{amsmath,amssymb}
\usepackage{graphicx}
\usepackage{amsfonts}
\begin{document}
\title{Short wavelength regenerative amplifier free electron lasers}
\author{D. J. Dunning}
\address{ASTeC, Daresbury Laboratory, Warrington WA4 4AD, UK}
\author{B. W. J. M$^{\mathrm{c}}$Neil and }
\address{SUPA, University of Strathclyde,Glasgow G4 0NG, UK}
\author{N. R. Thompson}
\address{ASTeC, Daresbury Laboratory, Warrington WA4 4AD, UK}
\maketitle

\begin{abstract}
In this paper we discuss extending the operating wavelength range of
tunable Regenerative Amplifier FELs to shorter wavelengths than
current design proposals, notably into the XUV regions of the
spectrum and beyond where the reflectivity of broadband optics is
very low. Simulation studies are presented which demonstrate the
development of good temporal coherence in generic systems with a
broadband radiation feedback of less than one part in ten thousand.
\end{abstract}

\section{\label{sec:level1}Introduction}
A regenerative amplifier free-electron laser (RAFEL) is a high-gain
resonator FEL which achieves saturation in a few round-trips of the
radiation in a high-loss, and hence low feedback, optical cavity.
Because the radiation feedback fraction is low it is feasible that
the use of low reflectivity optics in the resonator makes the RAFEL
a candidate for short wavelength operation \cite{mcneil:1990}.
Several RAFEL proposals have been made in the VUV \cite{CDR,NJP} and
X-ray \cite{huang} and some experimental results obtained
\cite{nguyen:2001, faatz}.

There are several expected advantages of the RAFEL over other types
of FEL. The RAFEL should be less sensitive to radiation-induced
mirror degradation than a low gain oscillator FEL, and the small
number of passes required to reach saturation should relax the
longitudinal alignment tolerances. The optical feedback also allows
the undulator length to be reduced compared to a Self Amplified
Spontaneous Emission (SASE) FEL, and it is expected that because of
the feedback a RAFEL source can deliver higher quality and more
stable pulses than a SASE FEL.

The properties of the transverse modes within the cavity differ from
those of a low-gain oscillator FEL. Because of the high loss of the
resonator the radiation is not stored over many passes, and because
of the high-gain of the FEL the radiation does not propagate freely
within the cavity but experiences gain guiding. The cavity's primary
function is to return a small field to the start of the undulator to
seed the interaction with the subsequent electron bunch. For these
reasons it is equally valid to refer to a RAFEL as a High-Gain
Self-Seeding Amplifier FEL.

In this paper we present 1D modeling results for a system with a
very low feedback factor that returns only $\sim 10^{-5}$ of the
undulator output. Such low feedback may occur when mirror
reflectivities are very poor, for example into the XUV and x-ray
regions of the spectrum. The results are encouraging and suggest
that, in principle, a low feedback RAFEL may prove a viable source
at these photon energies.

\section{A Generic High-Gain RAFEL}
\begin{figure*}[ht]
\begin{center}
\includegraphics[width=160mm]{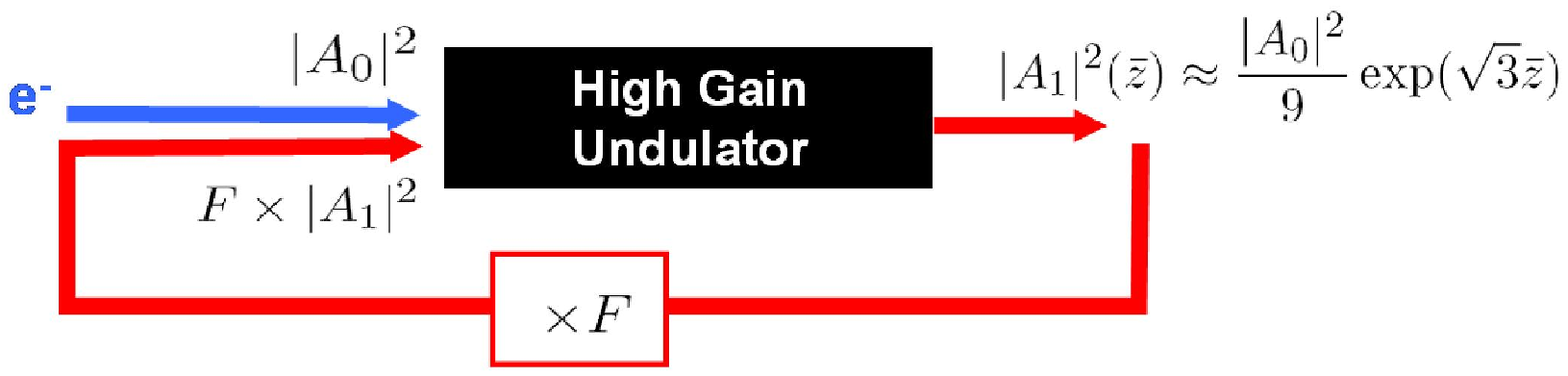}
\end{center}
\caption{Schematic representation of a generic high gain RAFEL
system.} \label{schematic}
\end{figure*}
We now consider  a generic high gain system shown schematically in
Fig.~\ref{schematic} and investigate the properties of such a system
when the feedback fraction is reduced to very low levels. First we
optimise the feedback fraction using two criteria---the output power
and the pulse temporal coherence should both be maximised. We work
in the units of the universal scaling \cite{BNP} where
$\bar{z}=z/l_g$ and $l_g=\lambda_w/4\pi\rho$ is the nominal gain
length, with $\lambda_w$ the undulator period and $\rho$ the FEL
parameter.

It can be shown from~\cite{kjk} that the electron beam equivalent
shot-noise power is:
\begin{equation}
|A_0|^2\approx \frac{6\sqrt{\pi}\rho}{N_\lambda\sqrt{\ln{\left (
N_\lambda / \rho \right)}}},
\end{equation}
where $N_\lambda$ is the number of electrons per radiation
wavelength. In the exponential gain regime the radiation intensity
after a single pass through an undulator of scaled interaction
length $\bar{z}$ is then given by
\begin{equation}
|A_1|^2(\bar{z})\approx \frac{|A_0|^2}{9}\exp(\sqrt{3}\bar{z})
\end{equation}
and after returning a fraction $F$ of the output power to the start
of the undulator, via some as yet undefined optical system, the seed
power at the start of the second pass is $F\times |A_1|^2$. The
necessary condition for the development of longitudinal coherence is
that this seed power is greater than the shot noise power, i.e.
$$
F\times |A_1|^2 > |A_0|^2.
$$
A feedback factor criterion to dominate the shot noise can then be
defined as:
\begin{equation}\label{F_N}
F_N > 9 \exp(-\sqrt{3}\bar{z}).
\end{equation}

\begin{figure}[ht]
\begin{center}
\includegraphics[width=90mm]{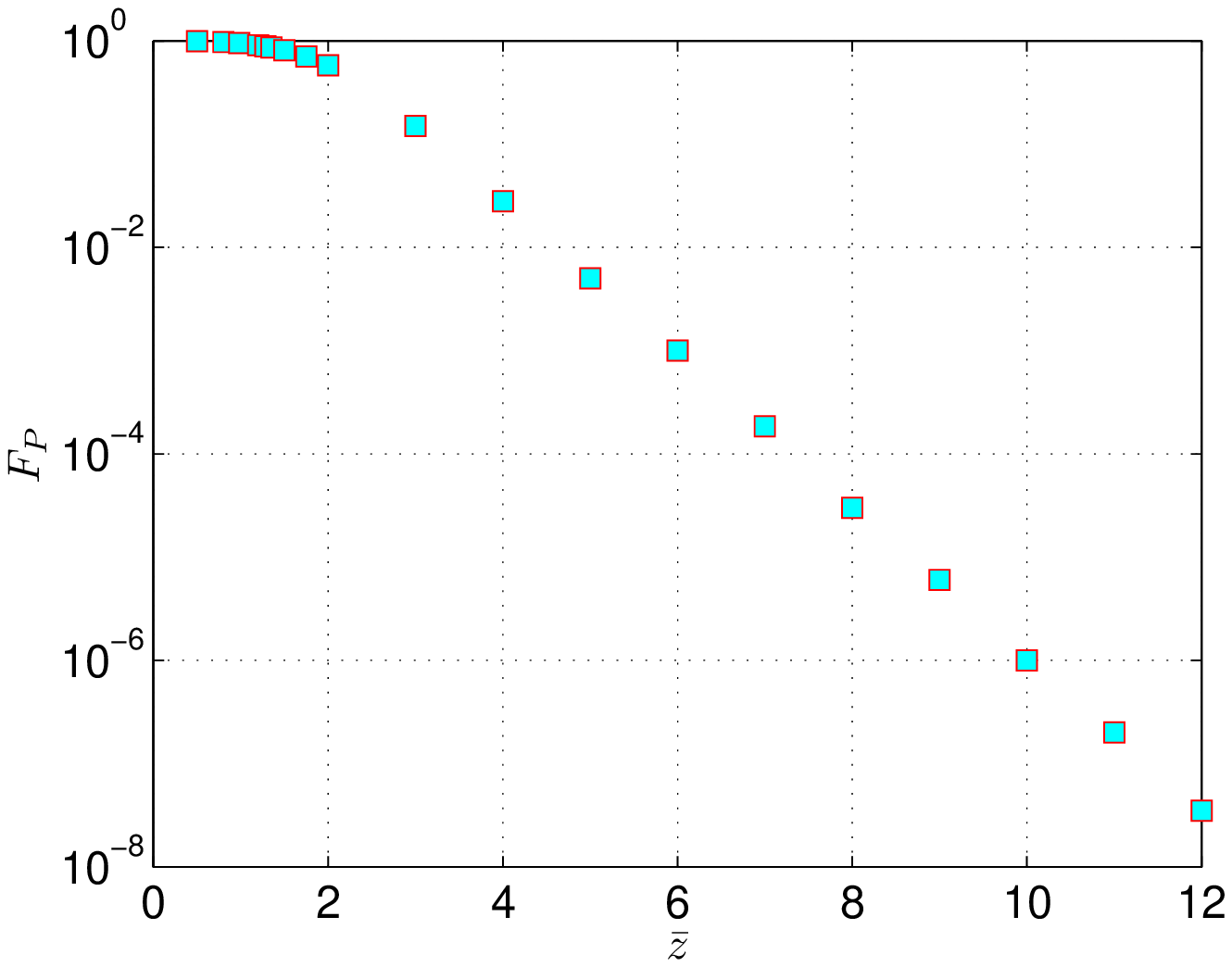}
\end{center}
\caption{Results of one-dimensional steady-state simulations to
determine the feedback factor $F_P$ that maximises the output power.
A fit to the numerical data over the range $3\leq \bar{z} \leq 12$,
gives $F_P \approx 25\exp (-\sqrt{3} \bar{z})$} \label{FIG:F_P}
\end{figure}
The feedback factor necessary to optimise the output power in the
steady state regime only has been determined from 1D simulations,
with the results shown in Fig.~\ref{FIG:F_P}. A fit to the numerical
data, valid over the range $3\leq \bar{z} \leq 12$, gives
\begin{equation}\label{F_P}
F_P\approx 25\exp (-\sqrt{3} \bar{z}).
\end{equation}
It is seen from comparison of (\ref{F_N}) and (\ref{F_P}) that
$F_P\simeq3F_N$ implying that optimising feedback to maximise the
output power will necessarily prove sufficient to dominate the
electron beam shot noise and enable the development of coherent
pulses. This postulate is tested with 1D time-dependent numerical
simulations in the next section.

\section{Time Dependent Simulations}
\subsection{Simulation Method and Parameters}
We choose a low feedback factor of $F_P=10^{-5}$ and use (\ref{F_P})
to derive the appropriate scaled interaction length of
$\bar{z}=8.67$. An FEL parameter of $\rho=2.9\times 10^{-3}$ is
used, typical of XUV FELs, with a peak number of electrons per
wavelength of $N_\lambda\approx 3.8\times 10^5$. The macroscopic
profile of the electron bunch is gaussian, and the input electron
beam is monoenergetic. The system is modelled using a 1D time
dependent code \texttt{FELO} \cite{FEL06VUV} which solves the 1D FEL
propagation equations
\begin{eqnarray*}
\frac{d\theta_j}{d\bar{z}} & = & p_j,\label{FEL1} \\
\frac{d p_j}{d\bar{z}} & = &
-(A(\bar{z}, \bar{z}_1)\exp[i\theta_j]+c.c.)\\
\left(\frac{\partial}{\partial\bar{z}}+
\frac{\partial}{\partial\bar{z}_1}\right)A(\bar{z}, \bar{z}_1)& = &
\chi (\bar{z}_1)\langle\exp[-i\theta]\rangle\equiv b(\bar{z},
\bar{z}_1) \label{FEL3}
\end{eqnarray*}
where $p$ is the particle energy $p=(\gamma-\gamma_r)/\rho \gamma$
with $\gamma_r$ the resonant electron energy in units of the
electron rest mass, $\theta$ the particle phase within the
ponderomotive well, $\bar{z}_1$ is the length along the electron
bunch from the bunch tail in units of the cooperation length
$l_c=\lambda_r/4 \pi \rho$ and  $\chi(\bar{z}_1)$ the function
describing the macroscopic electron current profile.

The feedback factor was varied from $F=10^{-3}$ to $F=2\times
10^{-6}$, and the cavity detuning value $\delta_c$, in units of
$\bar{z}_1$, varied from $\delta_c=0$, defining cavity synchronism,
to a detuned cavity length of $\delta_c=9.0$. For each combination
of these parameters the system was allowed to evolve over 200 cavity
round trips.

In order to compare the numerical results of the low feedback RAFEL
system with a SASE system, 200 separate simulations were done for an
equivalent SASE system with $\bar{z}=14$ where saturation of the
pulse energy is seen to occur.

\subsection{Simulation Results}
\begin{figure*}[ht]
\begin{center}
\includegraphics[width=185mm]{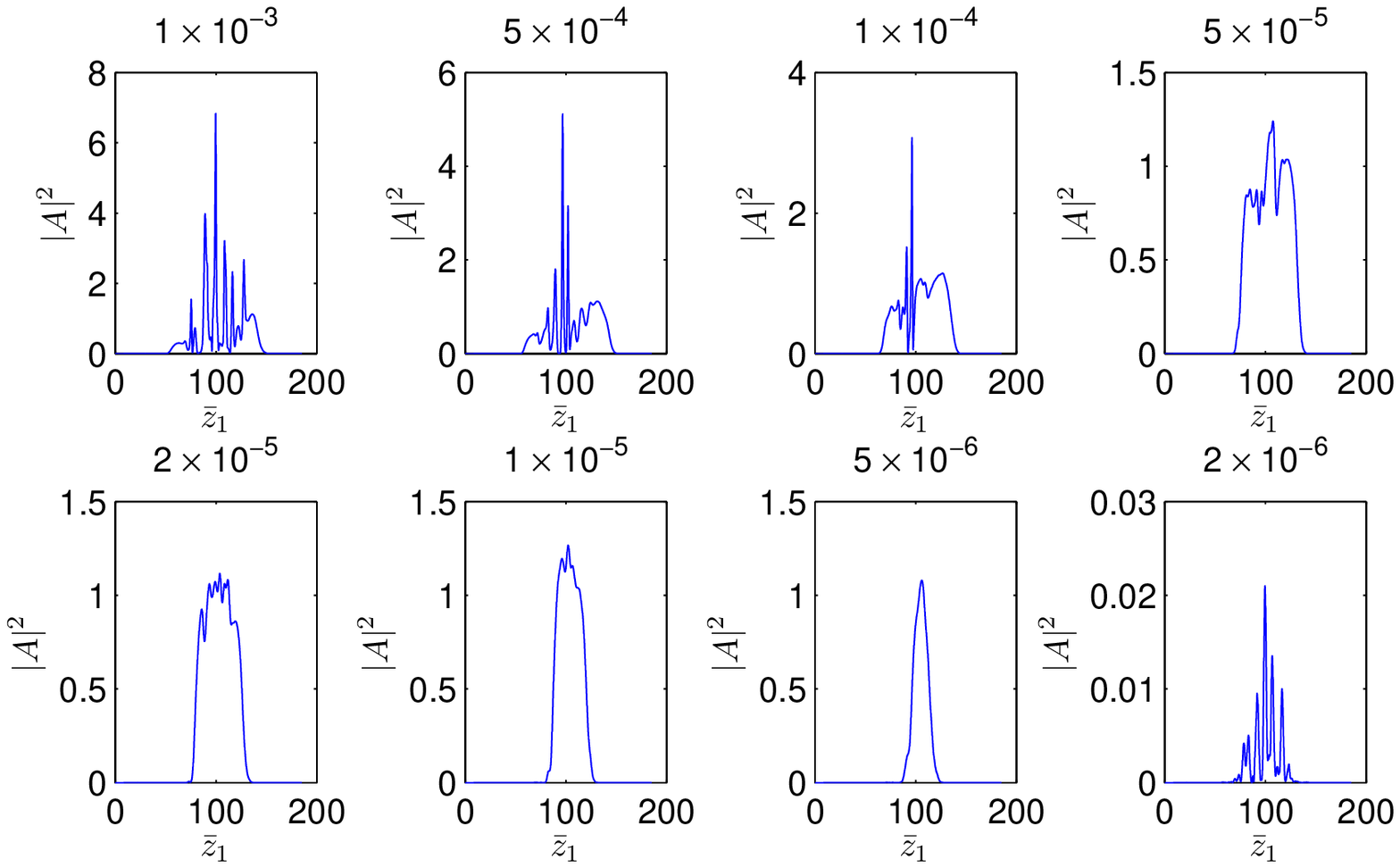}
\end{center}
\caption{Typical output pulses of the low feedback RAFEL system. The
feedback fraction is shown above each plot, and varies from
$F=1\times 10^{-3}$ (top left) to $F=2\times 10^{-6}$ (bottom
right). The cavity detuning value is $\delta_c=6.0$ for all pulses.}
\label{pulses}
\end{figure*}

\begin{figure*}
\begin{center}
\includegraphics[width=130mm]{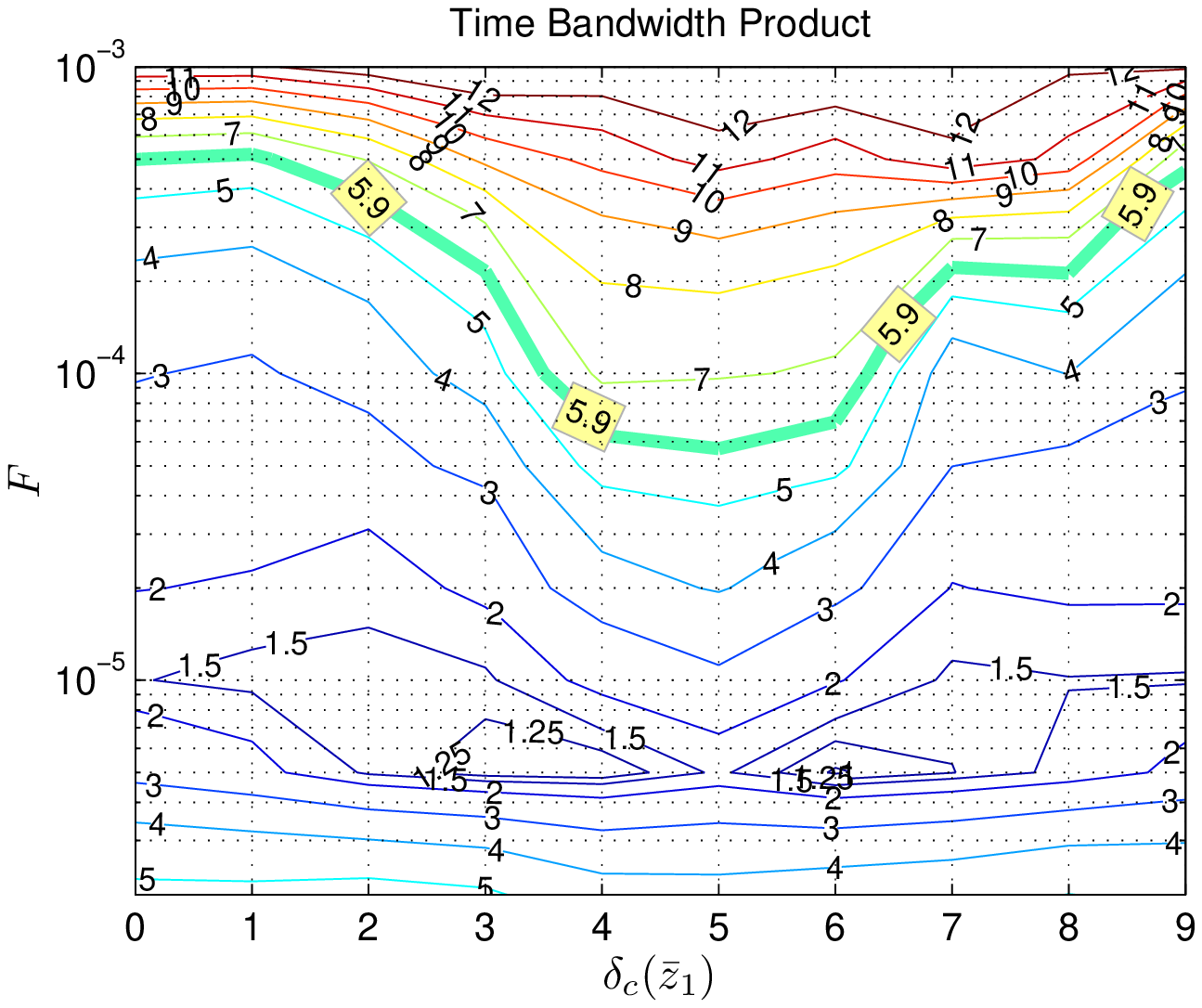}
\includegraphics[width=130mm]{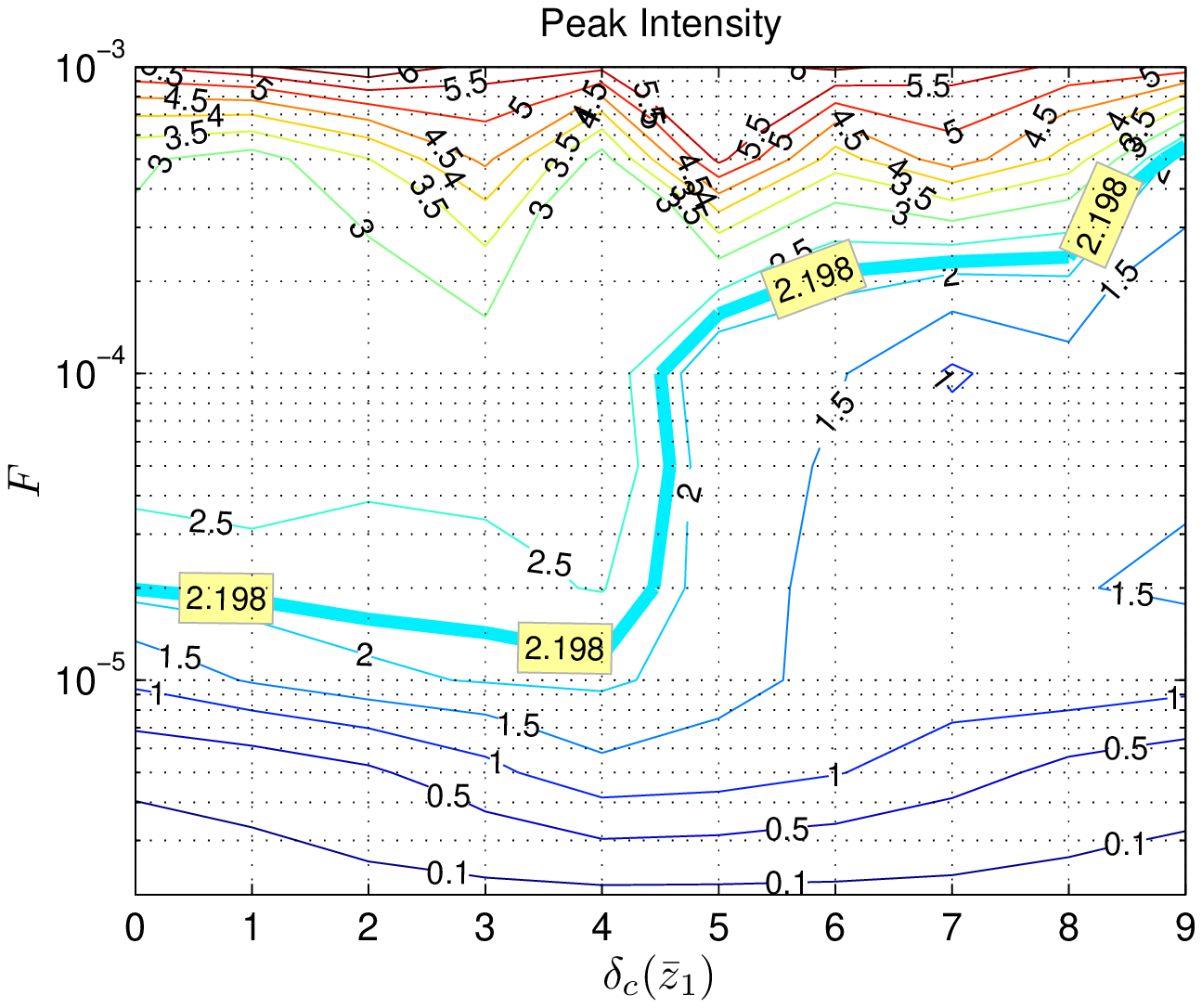}
\end{center}
\caption{The complete data for all simulations, in each case
averaged over 200 post-saturation pulses, for time-bandwidth product
$\langle\Delta \nu \Delta t\rangle$ (top), and peak intensity
$\langle|A|_{\mathrm{peak}}^2\rangle$ (bottom). In each plot the
vertical axis gives the feedback $F$ and the horizontal axis the
cavity detuning $\delta_c$. The bold contour represents the averaged
value seen for the 200 SASE simulations.} \label{Contours1}
\end{figure*}

\begin{figure*}
\begin{center}

\includegraphics[width=130mm]{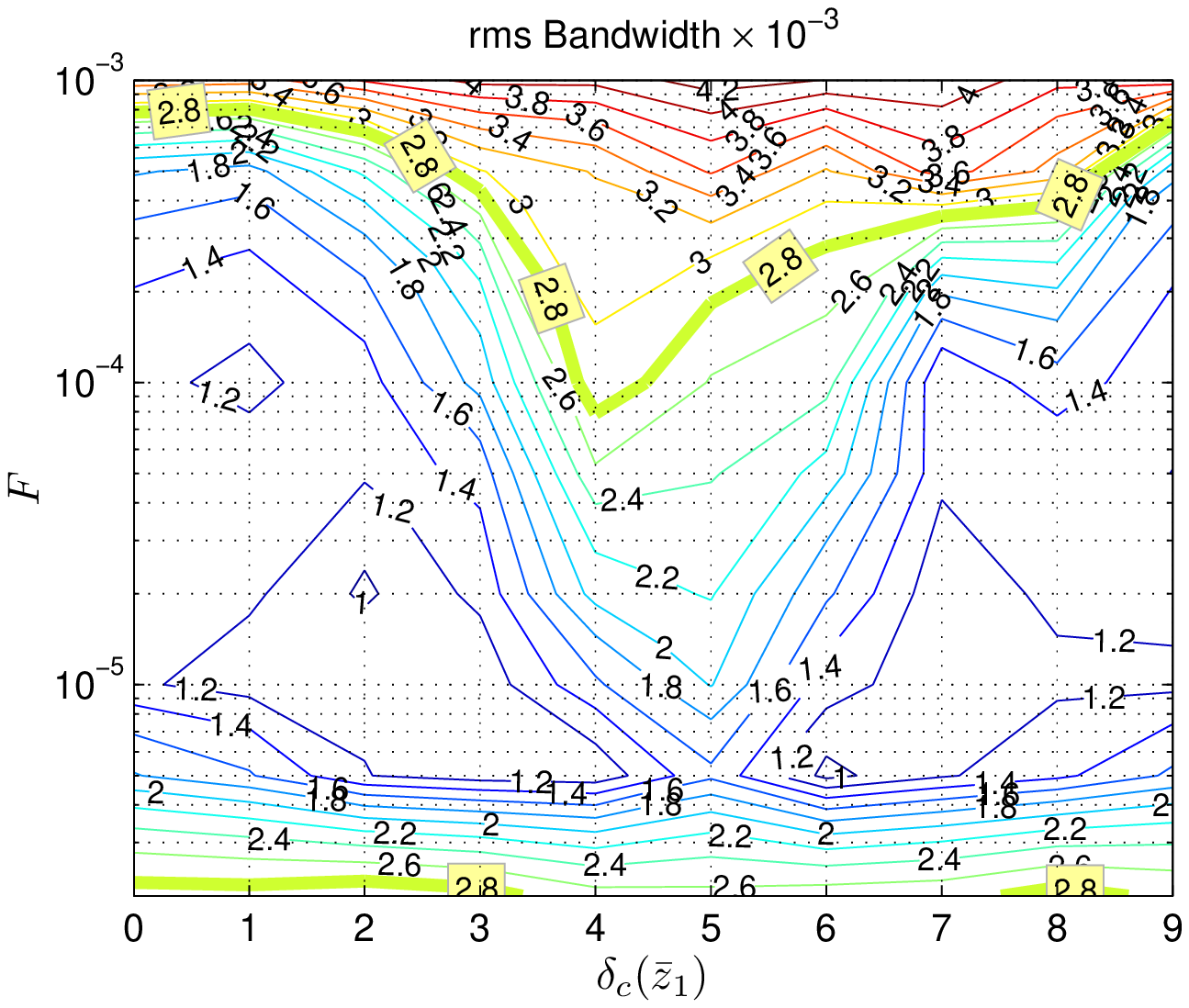}
\includegraphics[width=130mm]{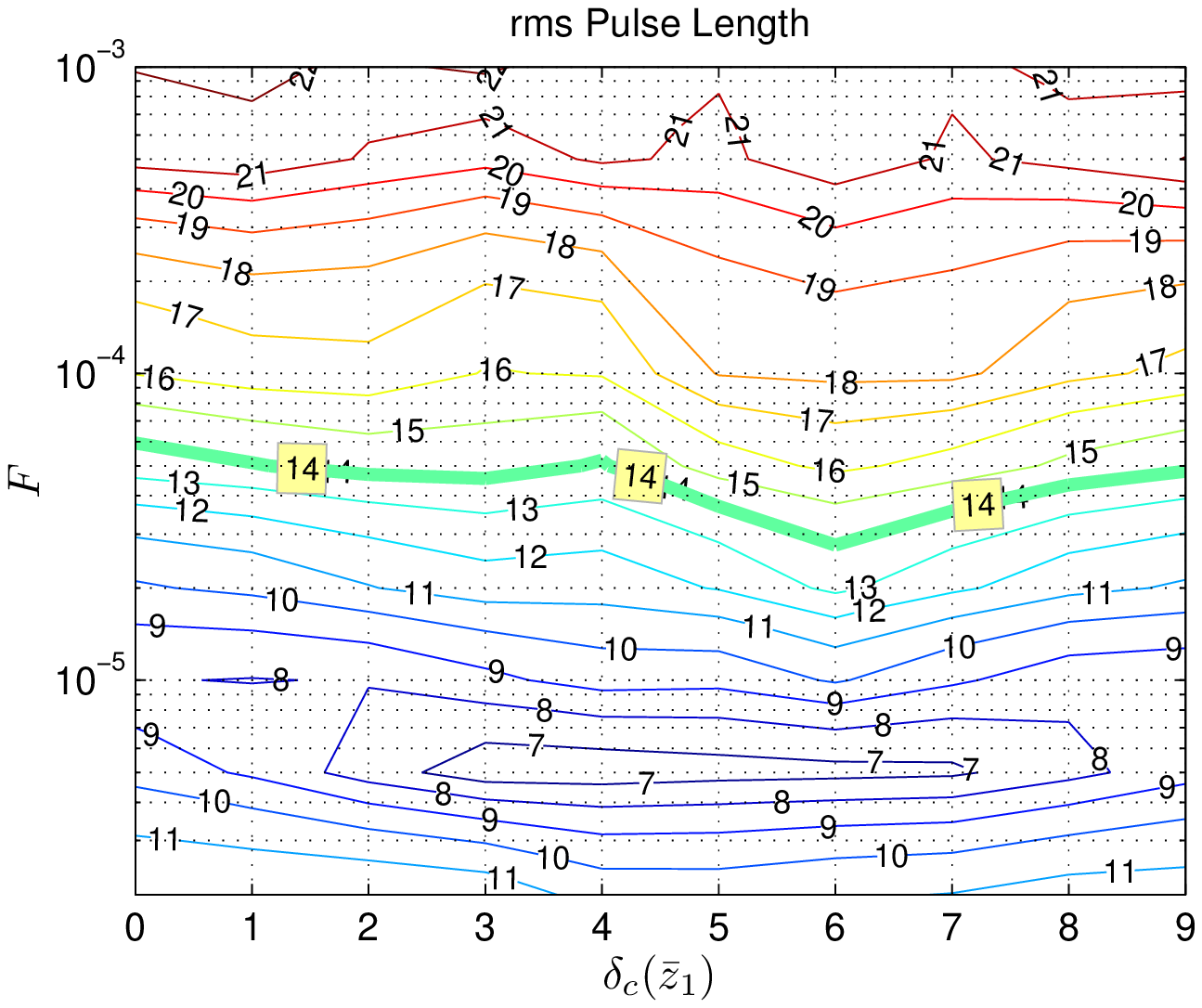}

\end{center}
\caption{The complete data for all simulations, in each case
averaged over 200 post-saturation pulses, for  rms bandwidth
$\langle\sigma_{\lambda}/\lambda\rangle$ (top) and rms pulse length
$\langle\sigma_{\bar{z}_1}\rangle$ (bottom). In each plot the
vertical axis gives the feedback $F$ and the horizontal axis the
cavity detuning $\delta_c$. The bold contour represents the averaged
value seen for the 200 SASE simulations.} \label{Contours2}
\end{figure*}

An analysis has been carried out to determine key parameters of the
output pulses pass-by-pass. The parameters of interest are the peak
intensity $|A|_{\mathrm{peak}}^2$, the rms pulse length
$\sigma_{\bar{z}_1}$, the rms relative linewidth
$\sigma_{\lambda}/\lambda$ and the time bandwidth product $\Delta
\nu \Delta t$ which is used to quantify the development of the
temporal coherence. The definition used is
\begin{equation}
\Delta \nu \Delta t = \frac{1}{\lambda}\left(\frac{\Delta
\lambda}{\lambda}\right)\Delta z \end{equation} with $\Delta z$ the
pulse width. The numerical value obtained depends on the definition
of width chosen. The choice used here here is $\Delta z = 2 \sqrt{2
\ln 2}\times\sigma_z$ under which definition a transform limited
gaussian intensity pulse would give the result obtained using FWHM
values of $\Delta \nu \Delta t \simeq 0.44$ (the relationship
between $\sigma$ and FWHM for a gaussian given by FWHM$(z) =  2
\sqrt{2 \ln 2}\times\sigma_z$).

\subsubsection{SASE results} The results of the SASE simulations are as follows: the root mean square (rms) linewidth
over 200 simulations was $\langle \sigma_{\lambda} / \lambda \rangle
=2.77 \times 10^{-3}$ with  an rms pulse length $\langle
\sigma_{\bar{z}_1}\rangle=14.01$ giving a time-bandwidth product of
$\langle \Delta \nu \Delta t \rangle=5.9$. The peak intensity
$\langle |A|_{\mathrm{peak}}^2 \rangle = 2.2$.

\subsubsection{Low feedback RAFEL Results}
To identify some features of the RAFEL output, pulse profiles for a
feedback fraction decreasing from $F=10^{-3}$ to $2 \times 10^{-6}$
are shown first in Fig.~\ref{pulses} for a cavity detuning value of
$\delta_c=6.0$. It is seen that for  $F=10^{-3}$ the pulse profile
is spiky with a peak intensity $|A|^2_{peak}=7$. The bandwidth for
these parameters, averaged over 200 post-saturation passes, is
$\langle\sigma_{\lambda}/\lambda\rangle=4\times 10^{-3}$, greater
than the mean SASE value of $\langle \sigma_{\lambda} / \lambda
\rangle =2.77 \times 10^{-3}$, and the time bandwidth product is
$\langle \Delta \nu \Delta t \rangle=12$ compared to the SASE value
of $\langle \Delta \nu \Delta t \rangle=5.9$. This data indicates
that the RAFEL pulse is over-saturated. The feedback fraction is too
high so that the seed power is too great and the RAFEL saturates
before the end of the undulator.

Fig.~\ref{pulses} shows that as the feedback fraction is decreased
the pulse profile becomes cleaner, with the front and back of the
pulse cleaning up first, leaving a spiky region in the centre of the
pulse. This behaviour is attributed to the gaussian electron current
profile---the front and back of the pulse experience less gain and
do not oversaturate whereas the centre of the pulse oversaturates.
The time bandwidth product falls below the SASE value at a feedback
fraction of $F=5\times10^{-5}$. For lower feedback, the time
bandwidth product continues to fall until it reaches a minimum value
of $\langle \Delta \nu \Delta t \rangle=1.0$ at a feedback of
$F=5\times 10^{-6}$. Examination of the pass-by-pass data shows
individual pulses with $\Delta \nu \Delta t =0.68$, close to that of
a transform limited gaussian pulse. Finally, as the feedback
fraction is reduced further to $F=2\times 10^{-6}$ it is seen that
there is insufficient feedback for growth to saturation, and the
pulse shown represents a pre-saturation SASE pulse for an
interaction length $\bar{z}=8.67$. This conclusion is supported by
the fact that the pulse parameters (except peak intensity) have
reverted back to close to their values for the SASE simulations.

The complete data for all simulations, in each case averaged over
200 post-saturation pulses, are summarised in the contour plots of
Fig.~\ref{Contours1} and Fig.~\ref{Contours2}. In each of these
plots the vertical axis gives the feedback $F$ and the horizontal
axis the cavity detuning $\delta_c$. The bold contour represents the
averaged value of the 200 SASE simulations so that, for example, in
the top left plot showing time-bandwidth product, the area below the
bold contour represents all those feedback and detuning combinations
in which the low feedback RAFEL pulses have a lower time-bandwidth
product, and hence improved temporal coherence, than the SASE case.

\subsection{Discussion of Results}

It is clear from these simulations that the feedback factor derived
in (\ref{F_P}), with a value $F_P=10^{-5}$ and for interaction
length $\bar{z}=8.67$, is sufficient to significantly improve the
temporal coherence of the output compared to SASE, over the full
range of cavity detuning values. The feedback corresponding to the
best temporal coherence is $F=5\times 10^{-6}$ which is a factor of
two larger than the criterion derived in (\ref{F_N}) required to
dominate shot noise which gives $F_N>2.7\times 10^{-6}$.

From examination of the contour plots in Fig.~\ref{Contours1} and
Fig.~\ref{Contours2}, and considering the previous discussions,
three broad regimes can be identified:
\begin{itemize}
\item For $F \gtrsim 10^{-4}$: the output has
 the characteristics of over-saturation; \
\item For $10^{-4}\gtrsim F \gtrsim 5 \times 10^{-6}$: the applied feedback improves the
pulse coherence over SASE;
\item For $F \lesssim 5\times 10^{-6}$: the feedback is insufficient
to give growth to saturation or improve the coherence, giving
unsaturated SASE output.
\end{itemize}


\section{Conclusion}

An overview of the properties of Regenerative Amplifier FELs has
been presented and a one-dimensional feasibility study of a generic
high-gain RAFEL system which functions using cavity feedback factors
as low as $5\times 10^{-6}$. It has been shown that such a system
may generate radiation pulses of greatly improved quality than that
possible using SASE. The greatest temporal coherence is seen when
the power feedback is approximately double the shot noise power.
Here the time bandwidth product, averaged over 200 pulses, is
$\langle \Delta \nu \Delta t\rangle \approx 1.0$, approximately
double that of a transform limited gaussian pulse. This is more than
five times better than the equivalent SASE result, with individual
pulses having a time bandwidth product as low as $\Delta \nu \Delta
t \approx 0.68$.

It is also seen that if the feedback factor is too high the pulses
oversaturate and their properties are similar to, or worse than, the
equivalent SASE case.

Methods of attaining the low feedback factors were not discussed,
however the fact that they may be so small indicates that there is
significant scope in extending the low feedback RAFEL concept into
the XUV and possibly further. The possibility of combining harmonic
generation methods~\cite{bonifaciotwowigg,lhysci,harlase,twobeam}
and RAFEL also exists and these exciting possibilities will be the
subject of future research.


\end{document}